\documentclass[]{aa}
\usepackage{natbib,graphics,graphicx,psfig,amssymb,amsmath}
\bibpunct{(}{)}{;}{a}{}{,}

\newcommand{\Lsun}{$L_{\odot}$}

\newcommand{\revised}{\rm}

\begin{document}

\title{Flaring and self-shadowed disks around Herbig Ae stars:
simulations for 10~$\mu$m interferometers}
\author{R.~van Boekel\inst{1,2} \and C.P.~Dullemond\inst{3} 
\and C.~Dominik\inst{1}}
\institute{
Sterrenkundig Instituut `Anton Pannekoek',
Kruislaan 403, 1098 SJ Amsterdam, The Netherlands 
\and 
European Southern Observatory,
Karl-Schwarzschild-Strasse 2, D-85748 Garching, Germany
\and 
Max-Planck-Institut f\"ur Astronomie Heidelberg, K\"onigstuhl 17,
Heidelberg, Germany
}
\date{DRAFT, \today}

\offprints{R. van Boekel: boekel@mpia.de}

%\date{received\dots, accepted\dots}
\date{Received $<$date$>$; accepted $<$date$>$}

\authorrunning{van Boekel et al.}
\titlerunning{Herbig Ae star disks with a 10~$\mu$m interferometer}

\abstract{We present simulations of the interferometric visibilities
of Herbig Ae star disks. We investigate whether interferometric
measurements in the 10\,$\mu$m atmospheric window are sensitive to
the presence of an increased scale
height at the inner disk edge, predicted by recent models.
Furthermore, we investigate whether
such measurements can discriminate between disks with a ``flaring''
geometry and disks with a ``flat'' geometry.  We show that both these
questions can be addressed, using measurements at a small number of
appropriately chosen baselines. 
The classification of Herbig Ae stars in two groups, based
on the appearance of the spectral energy distribution (SED), has been 
attributed to a difference in disk geometry. Sources
with a group\,I SED would have a flaring outer disk geometry, whereas
the disk of group\,II sources is proposed to be flat (or
``self-shadowed'').  We show that this hypothesis can be tested using
long-baseline interferometric measurements in the 10\,$\mu$m
atmospheric window.}

\maketitle

\section{Introduction}

Herbig Ae/Be stars (HAEBEs, Herbig~\citeyear{1960ApJS....4..337H}, for
a more recent review see Natta et al.~\citeyear{2000prpl.conf..559N})
are intermediate mass pre-main-sequence stars, surrounded by material
which is left from the star formation process.  A sub-group of mostly
late B and A-F type HAEBE stars (hereafter HAEs) show little or no
optical extinction, and usually have low mass accretion rates as
derived from radio analysis (Skinner et
al.~\citeyear{1993ApJS...87..217S}) and the lack of significant
veiling in optical spectra.  There is ample evidence that the
circumstellar material responsible for the large infrared excesses of
these stars is located in a circumstellar disk 
(e.g. Mannings \& Sargent~\citeyear{1997ApJ...490..792M}; Grady et
al.~\citeyear{2001AJ....122.3396G}; Augereau et
al.~\citeyear{2001A&A...365...78A};
Eisner et al.~\citeyear{2003ApJ...588..360E}).  Vink et
al.~(\citeyear{2002MNRAS.337..356V}) show that the gaseous component
has a disk-like geometry on scales of less than 0.1\,AU.

Whereas the presence of these circumstellar disks seems firmly
established, the structure of the disks is a matter of debate. Kenyon
\& Hartmann (\citeyear{1987ApJ...323..714K}) developed ``flaring''
disk models for T-Tauri stars, in which $H/R$ (the ratio of the disk
surface height to the distance to the star) increases with increasing
distance to the central star.  The flaring disk model was refined by
Chiang \& Goldreich (\citeyear{1997ApJ...490..368C}, hereafter CG97)
who introduced an optically thin surface layer responsible for the
infrared emission features generally seen in circumstellar disks.
Natta et al. (\citeyear{2001A&A...371..186N}) and Dullemond et
al. (\citeyear{2001ApJ...560..957D}) reconsider the CG97 model
in the context of HAe stars,
proposing that the innermost region of the disk has an increased scale
height: the \emph{``puffed-up inner rim''}. This configuration, which
results from hydrostatic equilibrium at the
directly irradiated inner rim,
naturally explains the near-infrared bump commonly observed in the
Spectral Energy Distribution (SED) of HAe systems (Natta et
al.~\citeyear{2001A&A...371..186N}).

Meeus et al.~(\citeyear{meeus_2001_haebe_iso}) noted that based on the
IR SED, HAEs can be divided into two main groups: ``group\,I'' sources
that have a very strong, rising IR excess peaking around 60\,$\mu$m,
and ``group\,II'' sources displaying a more moderate IR excess, lacking
the 60\,$\mu$m bump.  It was proposed that group\,I sources have a
``flaring'' geometry, allowing the disk to intercept and reprocess
stellar radiation out to large stellocentric radii.  In the outer
disk of group\,II sources, on the other hand, $H/R$ is approximately
constant, or decreasing with increasing distance to the star. The
inner disk shields the outer disk from direct irradiation by the
central star, hence the term ``self-shadowed'' disk.  This
substantially reduces
the amount of radiation absorbed locally, leading to
lower temperatures in the outer disk of a group\,II source.

Recent 2D modeling by Dullemond~(\citeyear{2002A&A...395..853D}),
Dullemond \& Dominik (\citeyear{2004A&A...417..159D}, henceforth DD04), has
quantitatively confirmed that both flaring and self-shadowed disks are
natural solutions of the equation of vertical hydrostatic
equilibrium in passive circumstellar disks (see also
section~\ref{sec:disk_model}).  These models form the basis of the
current study. There is ample evidence that group\,I and group\,II
disks indeed have a flaring and self-shadowed geometry, respectively
(e.g. Grady et al.~\citeyear{2004ApJ...608..809G}, 
Dullemond \& Dominik~\citeyear{2004A&A...421.1075D},
Leinert et al.~\citeyear{2004A&A...423..537L}). However, as this
is not an observational study, we will consistently refer to the models
as flaring/self-shadowed, rather than group~I/II (which is by
definition an SED classification).

With the advent of long ($\sim 10^2$\,m) baseline infrared interferometry using
large apertertures, it has now become possible to observe HAe disks in
the thermal infrared with a spatial resolution of order $10^{-2}$
arcsec.  At the present, the number of baselines will be limited, and only
interferometric amplitudes (no phases) are available. True
aperture synthetis imaging of disks is therefore not (yet)
possible. The interpretation of the measured visibility amplitudes, 
which contain information about the geometry of the disks, requires
the use of disk models.

{\revised In the near-infrared Herbig Ae/Be star disks have
been observed with long-baseline interferometers since a number of years
\citep{2001Natur.409.1012T,2001ApJ...546..358M,
2003ApJ...588..360E,2004ApJ...613.1049E,2005ApJ...623..952E}.  Up to recently
these measurements were evaluated using extremely simplified models: Gaussian
blobs, rings, ellipses etc. Such simple models made it possible to get a
handle on the typical size and inclination of the emitting source, but did not
go much further. For most sources the typical sizes were found to be in rough
agreement with those predicted by the inner rim models, but at high
accretion rates the observations deviate from predictions.  This is
explained by \cite{2005ApJ...622..440A} as due to the emission from accretion
inward of the inner dust rim, and by \cite{2005ApJ...624..832M} as due to the
protection of dust by optically thick gas, allowing the dust to survive closer
to the star. In these, and other, recent papers the modeling of the data
already starts to go well beyond the simple ring/ellipse models, using actual
multi-dimensional radiative transfer calculations in the case of Akeson et
al., and detailed accretion disk structure models in the case of
\cite{2003A&A...400..185L}. In particular with the new phase-closure
capabilities in the near-infrared at the IOTA and VLTI/AMBER interferometers
such more advanced models are clearly of great use.}

{\revised With the {\em mid-}infrared interferometry capabilities of the MIDI
instrument on the VLTI it is now possible to study the structure of the disk
at slightly larger scales than the inner rim. This is the region in which
the self-shadowed and flaring disks would most clearly be distinguishable.  A
first set of measurements was published by \cite{2004A&A...423..537L}, 
and a first tentative correlation between the shape of the disk
(flaring/self-shadowed) and the visibility was found. Mid-infrared
interferometry also has the interesting capability of measureming
mineralogical properties of the dust as a function of stellocentric
radius. First measurements of this kind  \citep{2004Natur.432..479V}
have revealed the strong radius-dependence of the crystallinity of dust, 
as predicted by theoretical models. In the present paper, however,
we will be mostly concerned with the first aspect of mid-infrared
interferometry: measuring the geometry of the disk.}

{\revised Based on the 2D disk models of \cite{2004A&A...417..159D}
as well as the simpler models of \cite{1997ApJ...490..368C} we
present calculations of the interferometric visibilities of HAe disks, to
investigate if it is possible to distinguish between the various disk 
geometries
predicted by these theoretical models. Since the mid-infrared probes the
structures at somewhat larger scale than the inner rim (from 1\,AU out to
about 20\,AU), this wavelength regime is more suited to our aims than the
near-infrared. The MIDI instrument is, so far, the only instrument capable
of doing such measurements for Herbig Ae/Be stars, so in our analysis we
focus on the typical baselines and properties of the VLTI.}

\begin{figure}[t]
  \centerline{
  \includegraphics[width=8.7cm]{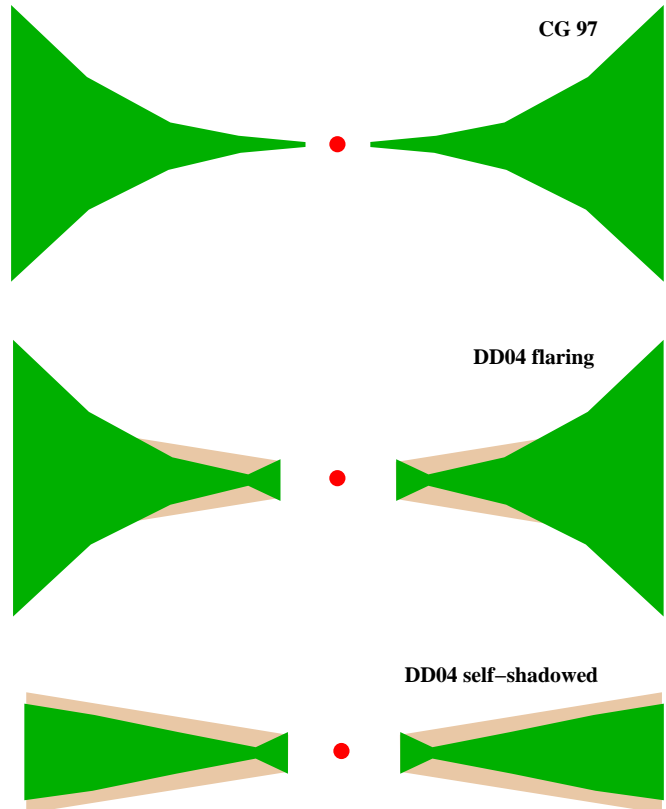}
  }
  \caption{A schematic representation of the geometries of the
disk models studied in this work. The shadow cast by the inner rim
is shaded light.
}
  \label{fig:pictograms}
\end{figure}

\section{Modeling method}
The goal of this study is to predict and compare interferometric
visibilities of various disk models. Synthetic disk images are made
using a ray tracing algorithm, where special care is taken to
ensure all spatial scales in the disk are sufficiently resolved.
Interferometric visibilities are calculated by Fourier transforming
the images. 

{\revised 
\subsection{The inner rim}
In the disk models considered here, the bright inner rim is treated
in a highly simplified fashion: it is a sharp, ``vertical wall''. As a
consequence, when such a model is viewed not pole-on, the flux from the near
side of the inner rim is strongly suppressed since the hot, irradiated rim
surface is occulted by the cooler parts immediately outside the rim surface.
The far side of the bright inner rim is in full view and is responsible for
essentially all of the near-infrared excess.  In this configuration, the total
amount of near-infrared emission depends strongly on the disk inclination,
suggesting that the observed strength of the ``3\,micron bump'' in the SED is a
measure of the latter. Observations of Herbig~Ae stars, however, show that
they all have rather similar near-infrared excesses, irrespective of their
inclination (Natta et~al. \citeyear{2000prpl.conf..559N}, Dominik
et~al. \citeyear{2003A&A...398..607D}).

This indicates that the appearance of the inner rim is more smooth
than the ``vertical wall'' used here. The processes that determine the shape
of the inner rim are currently not yet understood.  Isella \& Natta
(\citeyear{isella2005}) recently showed that the dependence of the evaporation
temperature on pressure naturally leads to a rounded-off inner rim.  When such
a disk is viewed at an inclination, both the near and the far side of the
inner rim will be bright (although still the far side will be brighter). The
bright inner rim will look like an inclined ring on the sky, rather than the
``half ring'' one obtains using the vertical wall model.

Realistic radiative transfer modeling of a rounded-off inner rim
introduces various numerical complications.  To avoid these difficulties we
adopt the simplified vertical rim structure used in Dullemond \& Dominik
(\citeyear{2004A&A...417..159D}). For a slightly off-polar inclination we
artificially circularize the disk emission to circumvent the
near-side/far-side asymmetry of the rim. In this way we mimic the rounded-off
shape of the rim without having to confront the numerical complexities of
radiative transfer in extreme optical depth rounded-off rims.  While the
spatial resolution of current 10~$\mu$m interferometers is just sufficient to
measure the diameter of the inner rim, observations at higher resolution are
required to study details of the rim structure. We therefore believe that
using this simplified approach is justified for our current purposes.}
To first approximation, inclination can be included by scaling the calculated
spatial frequencies (or interferometric baselines) by a factor 1/cos($\theta$)
along the minor axis of an inclined disk, where $\theta$ is the inclination of
the disk. At high inclinations, this approximation brakes down.

\subsection{Interferometry}
Rather than images, an interferometer produces an interference signal called
the interferometric ``visibility'' ($V$), which is the spatial
coherence function of the intensity distribution of the source.
The visibility is related to the intensity distribution through the 
van-Cittert-Zernike theorem, which states that the visibility is
the Fourier transform of the intensity distribution of the source. 
For an introduction to long-baseline interferometry, we refer to 
Lawson (\citeyear{2000plbs.conf.....L}). 

\begin{figure}[t]
  \centerline{
  \includegraphics[width=9.3cm]{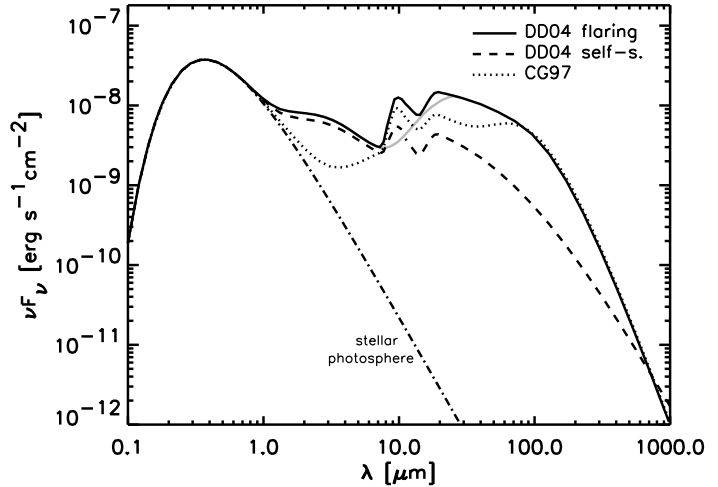}
  }
  \caption{The spectral energy distributions of the 
DD04 flaring (full curve), DD04 self-shadowed (dashed curve) and
CG97 (dotted curve) models. The full grey curve represents the 
DD04 flaring model, where the silicate resonances between 8 and 
25\,$\mu$m have artificially
been removed from the opacity table (see also 
section~\ref{sec:visibility_wavelength}
and Fig.~\ref{fig:run_B1_grey}).
}
  \label{fig:SEDs}
\end{figure}

\subsection{Spectrally resolved visibilities}
Traditionally, visibility curves are represented as
a function of the projected baseline $B$\footnote{The
spatial frequency of the observation is: $k=B/\lambda$, where
$\lambda$ is the observing wavelength. The units of $k$ are 
cycles/radian if $\lambda$ and $B$ have the same units}, 
at a specific wavelength (a $V(B)$ curve).
Such a curve represents a number of visibility measurements
at different projected baselines, which usually requires the use of
multiple telescope pairs and/or moving telescopes.

A new possibility in the 10\,$\mu$m region is the use
of spectral dispersion with large wavelength coverage. When using an
instrument that has this capability, one can obtain a whole
``visibility curve'' in one single measurement. Unlike the common $V(B)$
curve, the $V(\lambda)$ curve obtained this way holds many visibility
values at
only one baseline. The spatial resolution of the observation
($\approx B/\lambda$) changes by almost a factor of 2 between 7.5 and
14\,$\mu$m. Detailed modeling is required to interpret $V(\lambda$)
curves. Most HAe stars show a prominent emission band between 8 and
12\,$\mu$m, due to silicate dust. The shape of this emission band
varies strongly, depending on chemical composition, particle size and 
lattice structure of the silicate grains.
When simulating interferometric visibilities using disk models, 
one finds that the detailed shape of the visibility curve depends on the
opacities used, i.e. on the dust properties. These are different from
star to star, and vary within a disk as a function of distance to the 
star \citep{2004Natur.432..479V}. The visibilities measured in
the silicate emission feature are a mixture of disk structure and
mineralogy. Therefore, in order to deduce information on the disk
structure, it is preferable to use measurements at wavelengths
outside the silicate feature, which in practice means between
about 12 and 13.5\,$\mu$m. At 8\,$\mu$m, it is also possible to
sample the continuum emission, this is however more difficult since
here the atmospheric transmission is rather poor.

\begin{figure}[t]
  \centerline{
  \includegraphics[width=9.3cm]{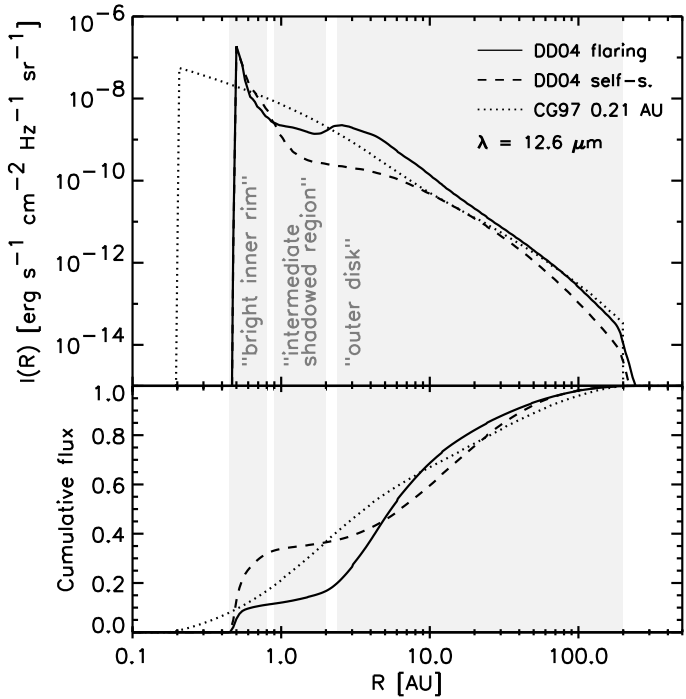}
  }
  \caption{Radial intensity profiles of the DD04 flaring, 
DD04 self-shadowed and
CG97 models (upper panel). The lower panel shows the normalized
cumulative flux distribution of the models. The three regions of the disk
that we distinguish (``bright inner rim'', ``intermediate shadowed region'',
and ``outer disk'') are indicated for the DD04 flaring model (in the
DD04 self-shadowed model, the intermediate shadowed region extends
somewhat further outward).
}
  \label{fig:group_I_II_radial}
\end{figure}

\subsection{The DD04 disk model}
\label{sec:disk_model}
The disk models used in this work are described in DD04.  These are
2D axisymmetric models in which the gas and dust density and
temperature are given as a function of radius $R$ and polar coordinate
$\Theta$. The disk is assumed to be heated only by irradiation by the
central star.  A 2D continuum radiative transfer code is used to
compute the entire temperature structure of the disk.  The vertical
density structure, for a given radial surface density distribution
$\Sigma(R)$, is computed by demanding vertical hydrostatic
equilibrium. In this way the disk has a self-consistent temperature
and density structure, from which images and SEDs can be computed
using a ray-tracer. For this work we use the following stellar
parameters: $M_{*}=2.5\,$M$_{\odot}$, $R_{*}=2\,$R$_{\odot}$ and
$T_{*}=10\,000$\,K, which amounts to a stellar luminosity of
$L_{*}=36\,$L$_{\odot}$.  All disk models in this work have a disk mass
of 0.1\,M$_{\odot}$, a gas-to-dust ratio of 100, a surface density
distribution of $\Sigma \propto R^{-1.5}$, and an outer radius of
200\,AU. The DD04 models have an inner disk radius which is
calculated self-consistently assuming an optically thick inner rim.
The location of the inner rim is set by the dust evaporation
temperature, which is about 1500\,K for silicate dust.  The CG97 model
has an inner radius of 0.21\,AU, which corresponds to the radius where
the black-body temperature is 1500\,K.
For the dust opacities we use a simple model consisting only
of small silicate grains (Laor \&
Draine~\citeyear{1993ApJ...402..441L}).

If the disk is optically thick enough, the disk has a flaring shape
(DD04 and Dullemond~\citeyear{2002A&A...395..853D}). When the optical
depth is decreased, a flaring disk can turn into a self-shadowed disk
and the SED changes from a group\,I to a group\,II shape. The flaring
and self shadowed models shown here are the BL1 and the BL4 model from
DD04. Both disks have a mass of 0.1\,M$_{\odot}$. In the BL1 model all
the dust mass is in 0.1\,$\mu$m silicate grains. In the BL4 model
99.9\% of the mass has been converted into 2\,mm size grains located
in the midplane, while
only 0.1\% remains in small 0.1\,$\mu$m grains, thus strongly lowering
the opacity of the disk. We stress however, that the flaring
vs. self-shadowed behaviour of the disk depends on high
vs. lower \emph{optical depth}, and that dust coagulation is a 
possible mechanism to achieve lower optical depths.

In the proposed scheme, the outer disk of a self-shadowed source is
shielded from direct stellar irradiation by its own inner disk.
However, the outer disk receives near-IR radiation emitted by the hot
innermost disk regions, and optical/UV radiation which is scattered by
the diffuse inner disk atmosphere.  Therefore, the temperature and
scale height in a self-shadowed disk are still significantly larger
than zero.  Note that in a flaring disk, there is also a region just
outward of the inner rim that is shielded from direct stellar
radiation. Contrary to a self-shadowed disk, a flaring disk emerges
from the shadow cast by the puffed up inner rim, at distances of a few
AU from the star.

\section{Results}
\label{sec:results}
The geometry of the disk models investigated in this work is
schematically represented in Fig.~\ref{fig:pictograms}.
The outer disk geometries in the CG97 and DD04 flaring models are very similar,
however the inner disk region is different. The DD04 flaring model has
a puffed up inner rim, and an intermediate shadowed region
(shaded light). The DD04 self-shadowed model has an inner disk
structure that is very similar to the DD04 flaring model.
The outer disk in the self-shadowed model never rises above
the shadow cast by the puffed up inner rim, but is irradiated by
the hot inner disk regions.

\subsection{Spectral energy distributions}
\label{sec:SEDs}
The emerging spectral energy distributions of the disk models are
shown in Fig.\,\ref{fig:SEDs}.  The infrared excess emission at far
infrared (FIR) wavelengths ($\sim$60-100\,$\mu$m) is clearly stronger
in the flaring disk models than in the self-shadowed model. The FIR
excess is significantly stronger in the DD04 flaring model than in the
CG97 (also flaring) model.  This can be traced to the simplifications
made in the CG97/DDN01 models which do not take properly into
account various 3D radiative transfer effects. In particular, these
models do not account for the moderate `boosting' of radiation toward
the polar axis to compensate for the occultation in equatorial
directions by the disk's own flaring outer regions. In the DD04 models
these effects are consistently taken into account by virtue of the
full multi-dimensional radiative transfer treatment used in those
models.  The near-infrared excess around 2-3\,$\mu$m which is very
prominent in the DD04 model SEDs, is much less pronounced in the CG97
model. In the 10\,$\mu$m region, the CG97 and DD04 flaring model SEDs
are virtually identical in spectral shape, though the DD04 flaring
model has a somewhat higher absolute flux level.  The DD04
self-shadowed disk is fainter than the flaring models, and has a
bluer continuum slope in the 10\,$\mu$m region.

\subsection{Radial intensity profiles}
\label{sec:radial_intensity_profiles}
Fig.\,\ref{fig:group_I_II_radial} shows the radial intensity profiles
at 12.6\,$\mu$m of a DD04 flaring and self-shadowed disk model, and the
CG97 model. Both the DD04
flaring and self-shadowed model essentially exhibit three regimes: 
\begin{itemize}
\item[1)] the ``bright puffed up inner rim'', that causes a ringlike emission, 
contributing mainly between 0.5 and 0.8\,AU from the star.
\item[2)] a region just behind the inner rim (as seen from the star), 
where the dust temperatures are much lower than in the inner rim.
From this ``intermediate shadowed region'', relatively little radiation emerges
(as can be seen in the cumulative flux distributions of the DD04 models
in the lower panel of Fig.\,\ref{fig:group_I_II_radial}, which are nearly
constant in this region).
\item[3)] the ``outer disk region'', whose main
flux contribution arises between 3 and 20\,AU from the central star.
In a flaring disk model, this outer disk region is directly irradiated
by the central star.
\end{itemize}
The ``bright puffed up inner rim'', ``intermediate shadowed region'' and 
``outer disk region'' are
of course just different parts of the same physical structure, and the
distinction made here serves merely to help the reader develop a
qualitative understanding of how such geometries translate into
interferometric visibilities.  
In Fig.\,\ref{fig:group_I_II_radial} we have indicated the three
regions discussed above. For the DD04 flaring model, the outer radius 
of the intermediate shadowed
region can be well defined to be between 2 and 3\,AU, where the
slope of the cumulative flux distribution clearly increases.
In the DD04 self-shadowed model, this radius is less clearly defined
but evidently somewhat larger than in the DD04 flaring model.
The contribution of the bright inner
rim emission to the total system flux depends strongly on wavelength
(for the self-shadowed model, the inner rim contributes more than 90\%
to the total flux at 6\,$\mu$m, about 60\% at 8\,$\mu$m, about 35\% at
13\,$\mu$m and less than
5\% at 30\,$\mu$m), and is always higher in a self-shadowed model than
in a flaring model harboring the same central star.  

The CG97 model has, per definition, a flaring disk structure. Contrary
to the DD04 flaring model however, it does \emph{not} have a bright
puffed up inner rim, and consequently it lacks an intermediate
shadowed region.

\begin{figure}[!t]
  \centerline{
    \includegraphics[width=9.3cm]{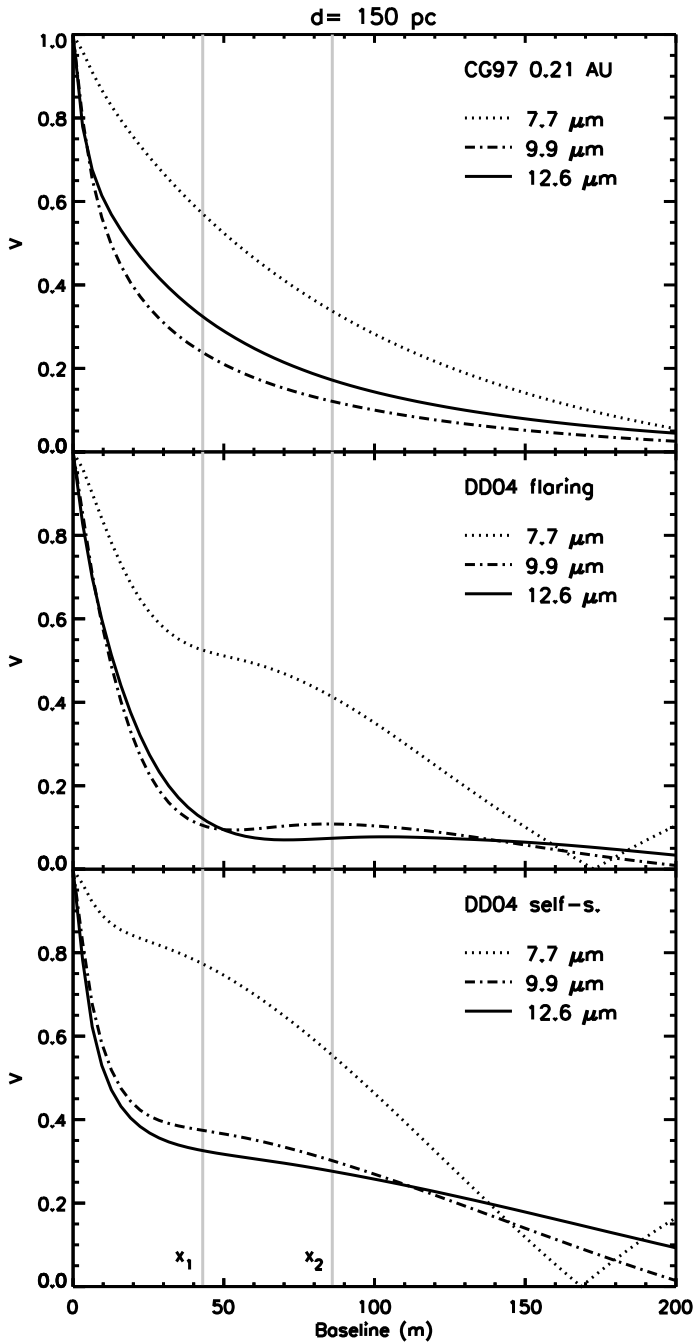}
  }
  \caption{Simulated visibility curves $V(B)$ of a CG97, 
a DD04 flaring, and a DD04 self-shadowed disk model (from top to bottom,
for $V(\lambda)$ curves of these models see 
Fig.\,\ref{fig:group_I_II_lambda2}).
We show model visibilities at three different wavelengths.
$x_1$ and $x_2$ denote optimum baselines to distinguish between the
various models (see section~\ref{sec:visibility_cc}).
}
  \label{fig:group_I_II_lambda1}
\end{figure}

\begin{figure}[!t]
  \centerline{
    \includegraphics[width=9.3cm]{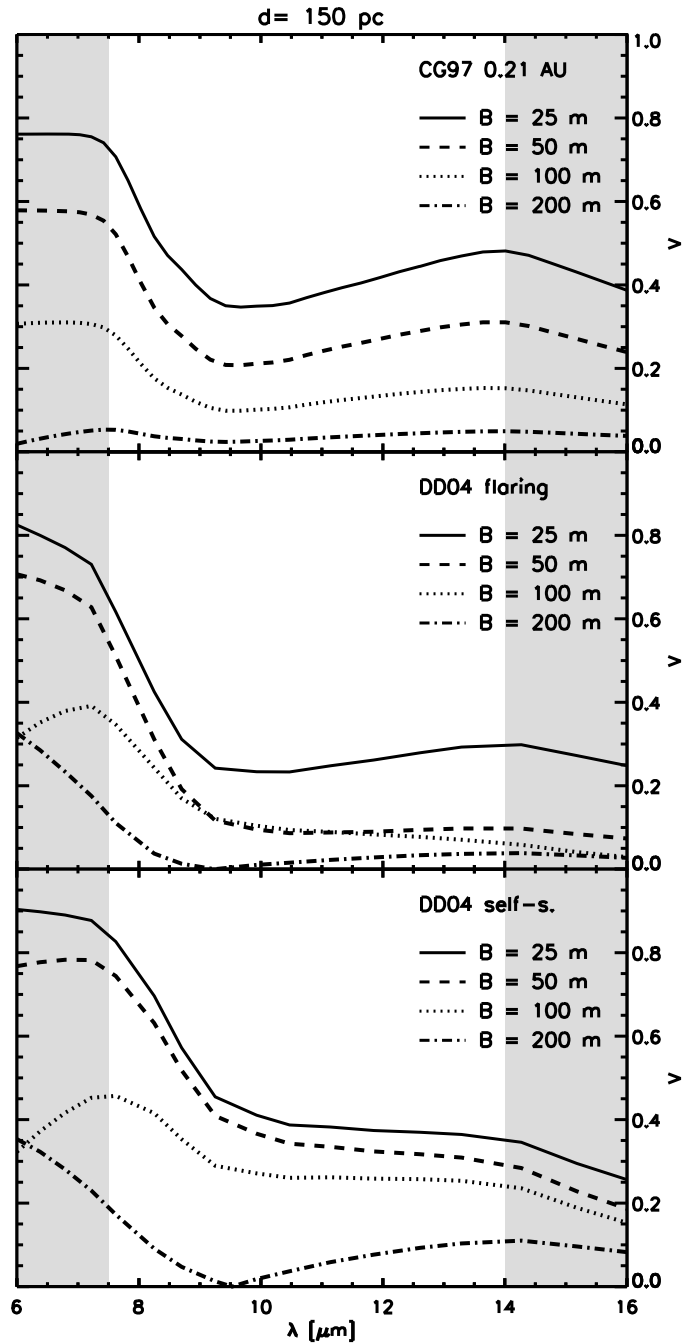}
  }
  \caption{Visibility curves $V(\lambda)$ of a CG97, 
a DD04 flaring, and a DD04 self-shadowed disk model (for $V(B)$ curves of
these models see Fig.\,\ref{fig:group_I_II_lambda1}).
We show predicted behaviour of the visibility as a function of
wavelength $V(\lambda)$, for several different baselines. 
The wavelength regions that are inaccessible from the ground are shaded grey.}
  \label{fig:group_I_II_lambda2}
\end{figure}

\subsection{Visibility curves}
\label{sec:visibility_curves}

\subsubsection{Visibility as a function of baseline}
\label{sec:visibility_baseline}
Fig.\,\ref{fig:group_I_II_lambda1} shows the predicted ``classical''
visibility curves $V(B)$ of the considered disk models, at several
wavelengths. From top to bottom we show the predictions for
the CG97 model, the DD04 flaring model and the DD04 self-shadowed model.
To develop an intuitive understanding of how the characteristics of the
emerging intensity distributions of the various models are reflected in
their visibility curves, we will discuss the curves in
Fig.\,\ref{fig:group_I_II_lambda1}. In this example,
the star is put at a distance of 150\,pc, typical for nearby Herbig stars.

At a baseline of 0\,m all sources are of course unresolved and have
a visibility of 1. The CG97 model shows a steady drop in visibility
as the baseline is increased. The slope of the visibility curve
changes gradually, reflecting that the radial intensity profile
shown in Fig.\,\ref{fig:group_I_II_radial} has no strong substructure.
In this sense, the CG97 model is ``scaleless''.

For the DD04 models, this is different. These models have essentially two
scales: the bright inner rim, which emits between 0.5 and 0.8\,AU, and
the outer disk, which emits most of its flux between about 3 and
20\,AU.  In between lies the intermediate shadowed region, whence little flux
emerges.  This general picture is reflected in the visibility curves.
Starting at 0\,m, and increasing
the baseline, we observe a steady drop in visibility as the outer disk
gets more and more resolved.  Note that at a baseline of 10\,m the 
visibility is already significantly
lower than 1, predicting that the largest modern day telescopes might
marginally resolve the outer disks in such objects at 10\,$\mu$m. For
the HAe star HD\,100546 this has indeed been observed 
(Liu et al. \citeyear{2003ApJ...598L.111L}, van Boekel et al. 
\citeyear{2004A&A...418..177V}).
At a baseline of about 30\,m, the outer disk is mostly resolved while
the inner rim is still essentially unresolved. Therefore, the
visibility curves flatten at this point. The visibility level at this
baseline (about 15\% for the flaring model and 40\% for the self-shadowed
disk, at 9.8\,$\mu$m) indicates the fraction of the total system flux
that is emitted by the bright inner rim.
At longer baselines, the bright inner rim itself becomes resolved by
the interferometer, and the visibility gradually goes to its first
null. Since the spatial resolution of the interferometer scales inversely with
wavelength, whereas the apparent diameter of the inner rim 
hardly depends on the wavelength, zero visibility is reached first at 
the shortest wavelengths, and at longer baselines for the longer wavelengths.

\subsubsection{Visibility as a function of wavelength}
\label{sec:visibility_wavelength}

Fig.\,\ref{fig:group_I_II_lambda2} shows $V(\lambda)$ curves for the
CG97, DD04 flaring and DD04 self-shadowed model, at a number of different
baselines. Each of these curves represents a \emph{single}, dispersed,
visibility measurement (the spectral region inaccessible from the
ground is shaded grey). Each curve in
Fig.\,\ref{fig:group_I_II_lambda2} can be regarded as a cut through
Fig.\,\ref{fig:group_I_II_lambda1} at a specific baseline, with a much
denser wavelength sampling.

{\revised
The overall trend for all curves is to show the highest visibilities at
8\,$\mu$m, and lower visibilities at 13\,$\mu$m.  This is because the apparent
size of the disks increases with wavelength more rapidly than the
interferometric resolution decreases.  There is generally a sharp decrease
in visibility between 8 and 10\,$\mu$m. There are two reasons for this.  For
the models with an inner rim one reason is that the emission from this rim
dominates the spectrum below about 8\,$\mu$m. The emission at 10\,$\mu$m
originates from more extended regions of the disk, resulting naturally in a
lower visibility than the 8\,$\mu$m emission. A second reason for the decline
of the visibility between 8 and 10\,$\mu$m -- and for the gentle rise in
visibility toward 13\,$\mu$m in some models -- is that the flux in the
10\,$\mu$m silicate feature originates predominantly from the warm surface
layers of the disk, while the flux outside the feature comes from the cooler
regions below. The warm dust in the surface layer can radiate in the
mid-infrared out to larger radii than the cooler dust in the disk interior.
In other words: in the warm surface layers the Wien exponential cut-off in
the mid-infrared takes effect at larger radii than in the disk interior.
This explanation also holds for the CG97 models, which do not have an inner
rim.}

\begin{figure}[t]
\hspace{-0.85cm}
    \includegraphics[width=9.8cm]{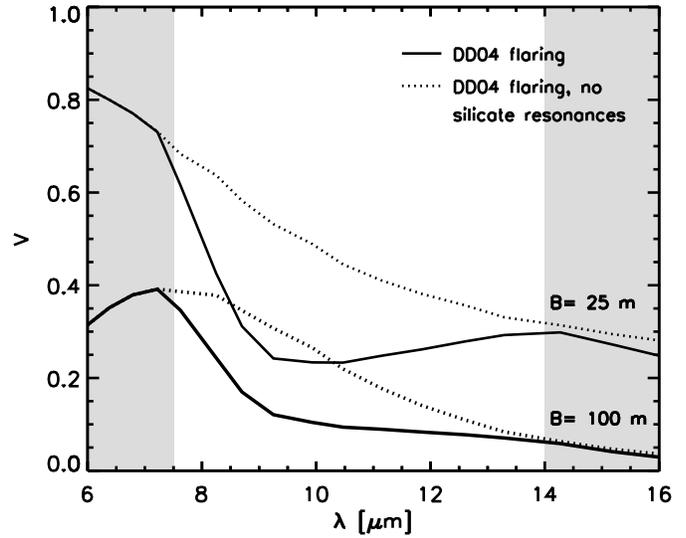}
  \caption{The effect of the increased dust opacity in the
silicate feature on the visibilities.
The full lines show the visibility curves of
the DD04 flaring model, the dashed lines show
the visibility curves of the same model, where the silicate
resonances have been removed in the opacities prior to the
ray tracing.
}
  \label{fig:run_B1_grey}
\end{figure}

In Fig~\ref{fig:run_B1_grey} we demonstrate the importance of the
silicate resonance for the simulated visibility curves. We show
the visibilities of the DD04 flaring model (full curves, see also the
middle panel of Fig.\,\ref{fig:group_I_II_lambda2}). To calculate
visibility curves of a model without silicate resonances, we removed
the 10 and 20 micron silicate features from the opacity table prior 
to the ray tracing (dotted curves).
The large influence of the opacity of the material on the resulting
visibilities is evident. The SED of the model without silicate 
resonances is indicated by the full grey curve in Fig.\,\ref{fig:SEDs}.

The interpretation of the curves in Fig.\,\ref{fig:group_I_II_lambda2}
in terms of disk geometry is
not straightforward, for several reasons. First, the emerging
intensity distribution of the disk changes with wavelength.
Second, the spatial resolution of the interferometer decreases
by almost a factor of two between the short and long wavelength
edges of the 10\,$\mu$m atmospheric window (N-band). 
Third, how the disk intensity distribution
(and thus the visibilities) changes with wavelength depends 
on the opacity of the dust, i.e. on mineralogy (Fig.\,\ref{fig:run_B1_grey}).
When we measure
a single $V(\lambda)$ curve, what we see therefore is a
mixture of disk geometry changing with wavelength, instrumental
resolution changing with wavelength, and the mineralogy of
the source.

As the mineralogy changes from star to star, it is difficult to obtain 
general diagnostics for disk \emph{geometry} from a $V(\lambda)$ curve.
The detailed interpretation
of such measurements requires a model of each individual star,
where both the spectrum (mineralogy) and the disk structure are
fitted simultaneously. 
As a general diagnostic for disk structure it is thus preferable to
measure visibilities in the continuum, where the visibilities do not
depend strongly on mineralogy. We will come back to this in
section~\ref{sec:visibility_cc}.

However, the spectral capabilities of the interferometer develop
their full strength, when one uses observations at several baselines
in order to reconstruct an ``image'' of the 
disk\footnote{In practice this will be easiest for disks that are seen not
too far from pole on, so their image has a high degree of azimuthal
symmetry.  This limits the number of baselines needed, and requires
the measurement of visibility amplitude only, since all phases will be
approximately 0}.  With spectrally dispersed visibilities, the
\emph{spectrum} of the disk is then known immediately at all positions
in the disk.  It is then possible to study the mineralogy, size
distribution and chemical composition of dust grains in the disk
surface layer as a function of distance to the central star, providing
crucial information about processing and radial mixing in disks.  

If the intensity distribution in the disk is strongly centrally peaked
like the models discussed in the present paper, the correlated flux
obtained at a single, long baseline\footnote{The visibility is the
ratio of the correlated flux and the total flux ($F_{\mathrm{cor}}/F_{\mathrm{tot}}$).
The correlated flux, or correlated spectrum in the case of a
spectrally dispersed measurement, is the quantity an interferometer
measures.}  can be directly interpreted as the spectrum of the
innermost regions of the disk (in the correlated spectrum obtained
with only one measurement, there is still an unknown spatial term
mixed in, that typically introduces a slope in the spectrum. This
however has little influence on the derived mineralogy).  The outer
disk spectrum can then be obtained as a difference between the
integrated disk spectrum and the inner disk spectrum.  Applying this
method to the first spectrally resolved full N-band visibility
measurements of HAe stars, it was demonstrated by \cite{2004Natur.432..479V}
that the mineralogy in the disk can vary strongly with distance to the star.

\begin{figure}[t]
\hspace{-0.85cm}
    \includegraphics[width=9.8cm]{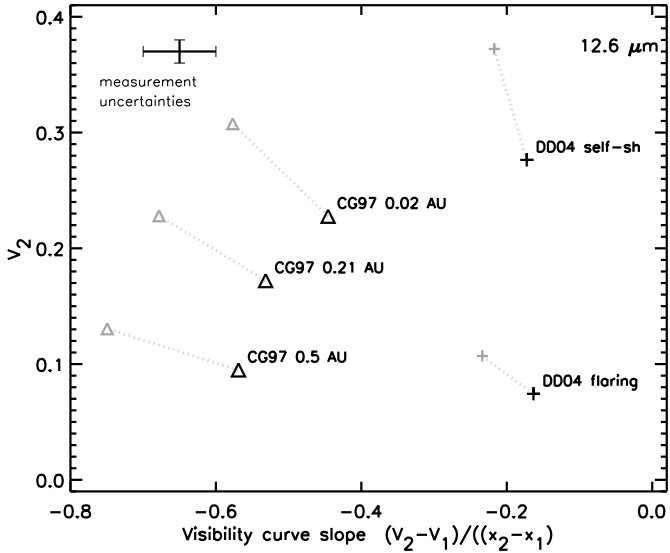}
  \caption{A diagnostic diagram used to distinguish between models
with (DD04) and without (CG97) a bright inner rim, and between models
with a flaring (DD04 flaring) and self-shadowed (DD04 self-sh) outer disk
geometry. On the horizontal axis we plot the slope of the visibility
curves between two appropriately chosen ``normalized'' baseline lengths
$x_1$ and $x_2$ ($x_i=B_i/d$, where $B_i$ is the baseline in m and 
$d$ is the distance to the star in pc, see section~\ref{sec:visibility_cc}
for how $x_1$ and $x_2$ are best chosen). On the vertical axis we plot 
the predicted visibility at the  longest baseline. The CG97 models
have a much steeper slope than the DD04 models. The DD04 self-shadowed
model has a much higher visibility than the DD04 flaring model.
{\revised In the upper left corner we have indicated the uncertainties due to
the limited precision of the visibility measurements, where we have
assumed a 1\% accuracy in visibility. In grey symbols we have plotted
where in the diagram the various models end up if we artificially
remove the silicate feature (as we did in Fig.\,~\ref{fig:run_B1_grey}).
This can be regarded as an upper limit for the uncertainty due to
mineralogy.}
}
  \label{fig:visibility_cc}
\end{figure}

\subsection{Distinguishing between the various models}
\label{sec:visibility_cc}
The goal of this study is to show how interferometric measurements
can be used to distinguish between the various disk models. Clearly,
the curves in Fig.\,\ref{fig:group_I_II_lambda1} are different
for the different models. However, one will typically not have continuous
measured visibility curves at hand, but rather have samples at
a few different baselines.
Here we show that it is, at least in principle, possible to distinguish
both between the CG97 and DD04 models on one hand, and between the
DD04 flaring and DD04 self-shadowed on the other, using measurements
at only two appropriately chosen baselines.

The distinction between the CG97 and DD04 models is based on the
absence of an intermediate shadowed region in the former. At spatial
scales corresponding to the intermediate shadowed region in the DD04
models, little flux emerges.  Therefore, the visibility curves are
relatively flat at the baselines corresponding to these spatial
scales, they show a ``plateau'' (very prominent in the DD04 flaring
model visibility at 12.6\,$\mu$m in
Fig.\,\ref{fig:group_I_II_lambda1}). The CG97 model does not have such
specific spatial scales with much reduced emergent intensity, and
therefore lacks the plateau in the visibility curve. The visibility
curves of CG97 and DD04 models thus have a different \emph{slope} at
baselines corresponding to the scale of the intermediate shadowed
region. Once this slope difference has been detected, DD04 flaring and
DD04 self-shadowded disks can be distinguished by the relative
contribution of the bright inner rim to the total system flux, which
is much higher for a self-shadowed model. Note that the bright inner
rim itself is virtually identical in both models, but the outer disk
is much brighter in the flaring disk than in the self-shadowed
case. Therefore, the predicted visibilities at our selected baselines
are much lower for the flaring model. We recall that deducing
properties about the disk structure is best done outside the silicate
emission feature, which in practice favours the region between 12 and
13\,$\mu$m.

A measurement at a specific baseline samples the corresponding
angular scale, and the physical scale (in AU) associated with this
baseline therefore depends linearly on the distance to the star.
It is therefore convenient to introduce the ``normalized baseline''
\begin{equation}
x\equiv\frac{B}{D}
\label{eq:normalized_baseline}
\end{equation}
where $B$ is the baseline in meter and $D$ is the distance to the star
in parsec. Consider the emission of a small part of the disk, arising
between $R$ and $R+\mathrm{d}R$ from the central star, i.e. a ring with angular
diameter $\theta=R/d$ (where $d$ is the distance to the star).
The visibility curve
of such an \emph{annulus} of emission is the zeroth order Bessel function:
\begin{equation}
V_{\mathrm{an}}(B)=J_0(\frac{\theta B}{\lambda})
\label{eq:ring_visibility}
\end{equation}
where $B$ is the interferometric baseline,
and $\lambda$ is the wavelength of
observation. The visibility reaches the first null at a baseline
of about
\begin{equation}
B_0\approx 158 \frac{\lambda_{\mu \mathrm{m}}}{\theta_{\mathrm{mas}}} \ [\mathrm{m}]
\label{eq:ring_B0}
\end{equation}
where for convenience the wavelength and annulus angular diameter have
been expressed in $\mu$m and milli-arcseconds, respectively.  Our goal
is to detect the effect of the intermediate shadowed region, i.e. the
very low flux contribution from annuli between about 0.8 and 3\,AU.  To
estimate which baseline is most sensitive to emission from an annulus
with diameter $\theta$, let us take the baseline where the visibility
has half its maximum value, $V_{\mathrm{an}}=0.5$:
\begin{equation}
B_{0.5}\approx 100 \frac{\lambda_{\mu \mathrm{m}}}{\theta_{\mathrm{mas}}} \ [\mathrm{m}]
\label{eq:ring_B0p5}
\end{equation}
For an annulus of 2\,AU radius, this corresponds to a normalized
baseline of $x=\lambda_{\mu \mathrm{m}}/40$. Let us take the region between 1 and
2\,AU from the star as characteristic for the intermediate shadowed region.

We find that in order to most clearly separate the specific DD04
models (with an intermediate shadowed region) used in this work from
the CG97 models (without an intermediate shadowed region), the best
choice for $x$ is:

\begin{equation}
\begin{array}{rcl}
x_1 &=& 0.038 \left(\normalsize{\frac{\lambda}{10 \ \mu \mathrm{m}}}\right) \left(\frac{L_{\ast}}{L_{\odot}}\right)^{\frac{1}{2}} \ \ \left[\frac{\mathrm{m}}{\mathrm{pc}}\right] \\
x_2 &=& 2x_1 \\
\label{eq:optimum_baseline}
\end{array}
\end{equation}

{\revised To allow the above formula to be applied to stars
of different luminosity we have applied a simple scaling of $x$ with
the square root of the luminosity.}
The visibility is now sampled at the baselines corresponding to $x_1$
and $x_2$ ({\revised for our 36\,\Lsun} star at 150\,pc and at a 
wavelength of 12.6\,$\mu$m these are 43 and 86\,m, respectively, see also
Fig~\ref{fig:group_I_II_lambda1}), yielding visibilities $V_1$ and
$V_2$.  In Fig.\,\ref{fig:visibility_cc} we plot the on the vertical
axis the visibility measured on the longest of the two baselines
($V_2$). On the horizontal axis we plot the deduced slope of the
visibility curve between $x_1$ and $x_2$.  In addition to the CG97
model with an inner radius of 0.21\,AU we plot CG97 models with inner
radii of 0.02 and 0.5\,AU, to illustrate the behaviour with varying
inner radius. From this figure we see that:
\begin{itemize}
\item[1)] The CG97 models have a much steeper slope than the DD04 models
in this baseline regime.
\item[2)] The DD04 flaring model has a much lower visibility at the
long baseline than the DD04 self-shadowed model (and on the short
baseline as well).
\end{itemize}
We can conclude therefore, that it is possible to discriminate between
the various models, with a very limited number of measurements.
{\revised In the upper left corner of Fig.\,\ref{fig:visibility_cc} we indicate
the uncertainty in the measured location of a source in this diagram due
to limited precision of the visibility measurments. We have here assumed
a visibility accuracy of 1\,\%, which is what MIDI aims to achieve.
The difference in the position of the models with and without a silicate
feature (plotted with black and grey symbols, respectively)
can be regarded as the extreme case for the uncertainty arising
from mineralogy, since for each individual source, we \emph{know} what 
the silicate feature looks like.}

This analysis has been done in an idealized world where we have both
assumed that the disks are pole-on, and that our models are a good
representation of the true disk geometry.
The optimum choice for $x_1$ and $x_2$ depends
on the geometry of the disk and may therefore in reality be somewhat
different than the values given in our recipe
(equation~\ref{eq:optimum_baseline}).  If the outer disk is smaller
than we predict, the value of $x_1$ should be increased. If 
the bright inner rim is located at radii somewhat
larger than predicted, the value of $x_2$ should be decreased. There is
however evidence that the bright inner rim is located at radii somewhat
smaller than predicted in our models (Eisner et
al.~\citeyear{2003ApJ...588..360E}), and therefore this is not likely
a reason for concern.
In practice, measurements at more than two baselines ($\approx$5) are
probably needed to unambiguously establish the nature of the sources.

\section{Conclusions}
We have presented model calculations of interferometric visibilities
of circumstellar disks around Herbig Ae stars. We compare predictions
for disks with (DD04) and without a bright inner rim (CG97). We show that
it is possible to distinguish between both model possibilities using
a small number of interferometric measurements in the 10\,$\mu$m
atmospheric window. Such measurements also allow to distinguish
between flaring and self-shadowed disk models. This
allows testing of the hypothesis that group\,I and group\,II sources
correspond to flaring and self-shadowed disks, as is
suggested by their spectral energy distributions.

\begin{acknowledgements}
R.~Lachaume and L.\,B.\,F.\,M.~Waters are gratefully acknowleged for
constructive comments on the manuscript.
\end{acknowledgements}

\bibliographystyle{aa}
\bibliography{2252}

\end{document}